\documentclass[preprint,prl,aps]{revtex4}

\begin{document}

\newcommand{\be}{\begin{equation}}
\newcommand{\ee}{\end{equation}}
\newcommand{\ben}{\begin{eqnarray}}
\newcommand{\een}{\end{eqnarray}}
\newcommand{\n}{\nonumber  }
\newcommand{\nn}{\nonumber \\ }
\newcommand{\nd}{\noindent}
\newcommand{\p}{\partial}

\title{Special features of the relation between Fisher
Information and Schr\"odinger eigenvalue equation}

\author{S.P. Flego$^1$}
\author{A. Plastino$^{2}$}
\author{A.R. Plastino$^{3,\,4}$}

\affiliation{ $^{1}$Universidad Nacional de La Plata, Fac. de Ingenier\'{\i}a,
 1900 La Plata, Argentina \\  $^{2}$Universidad Nacional de La Plata, Instituto
de F\'{\i}sica (IFLP-CCT-CONICET), C.C. 727, 1900 La Plata, Argentina \\
$^{3}$CREG-Universidad Nacional de La Plata-CONICET, C.C. 727, 1900 La Plata, Argentina\\
$^{4}$Instituto Carlos I de Fisica Teorica y Computacional and
Departamento de Fisica Atomica, Molecular y Nuclear, Universidad
de Granada, Granada, Spain}
%

\begin{abstract}
\nd \nd It is well known that a suggestive  relation exists that
links Schr\"odinger's  equation (SE) to the information-optimizing
principle based on Fisher's  information measure (FIM). The
connection entails the existence of a Legendre transform structure
underlying the SE. Here we show that appeal to this structure
leads to a first order differential equation for the SE's eigenvalues
that, in certain cases, can be used to obtain the eigenvalues
without explicitly solving SE. Complying with the above mentioned equation
constitutes a necessary condition to be satisfied by an energy eigenvalue.
 We show that the general solution is unique.

\vspace{2.cm}


\end{abstract}


\maketitle

\section{1. Introduction}

\nd It is well-known that a strong link exists between  Fisher'
information measure (FIM) $I$ and Schr\"odinger wave equation
(SWE) \cite{pla7,flego,reginatto,Univ,nuestro2}. In a nutshell,
this connection is based upon the fact that the constrained
minimization of $I$ leads to a SWE
\cite{pla7,flego,reginatto,Univ,nuestro2}. This, in turn, implies
intriguing relationships between various aspects of SWE, on the
one hand, and the formalism of statistical mechanics as derived
from Jaynes's maximum entropy principle, on the other one. In
particular, fundamental consequences of the SWE, such as the
Hellmann-Feynman and Virial theorems, can be re-interpreted in
terms of a special kind of reciprocity relations between relevant
physical quantities similar to the ones exhibited by the
thermodynamics' formalism \cite{Univ,nuestro2}. This demonstrates
that a Legendre-transform structure underlies the non-relativistic
Schr\"odinger equation. In this communication we show that such
structure allows one to obtain a first-order differential equation
 that energy eigenvalues must necessarily satisfy.

\section{2. Basic ideas}
\nd Fisher Information measure has been successfully applied to
the study of several physical scenarios, particularly quantum
mechanical ones (as a non-exhaustive recent set, see for instance
\cite{olivares,frieden3, KSC10,FS09,U09,LADY08,SAA07,N07,N06}). We
will briefly review here the pertinent formalism.
 \nd If an observer were to make a
  measurement of   $x$ and
  had to best infer $\theta$ from such  measurement,   calling the
    resulting estimate $\tilde \theta=\tilde \theta(x)$, one might
       wonder how well $\theta$ could be determined.
       Estimation theory~\cite{frieden2}
   asserts that the {\it best possible estimator} $\tilde
   \theta( x)$, after a very large number of $x$-samples
  is examined, suffers a mean-square error $e^2$ from $\theta$
  obeying the rule $Ie^2=1$,
  where the Fisher information measure (FIM) $I$, a functional of the PDF, reads
 \be \label{eq.1-1} I \,=\,\int ~dx ~f(x,\theta)
\left\{\frac{\partial ~ }{\partial
\theta}~\ln{[f(x,\theta)]}\right\}^2.\ee Any other estimator must
have a larger mean-square error (all estimators must be unbiased,
  i.e., satisfy $ \langle \tilde \theta({\bf x}) \rangle=\,\theta
  \label{unbias}$).   Thus, FIM has a lower bound. No matter what the parameter $\theta$ of the system might
  be,    $I$ has to obey
\be \label{rao} I\,e_\theta^2\,\ge \,1,\ee the  celebrated
Cramer--Rao bound \cite{frieden2}. The particular instance of
translational families merits special consideration. These are
mono-parametric distribution families of the form
$f(x,\theta)=f(x-\theta),$ known up to the shift parameter
$\theta$. All family members exhibit identical shape. After
introducing the amplitudes $\psi$ such that the probability
distribution function (PDF) are expressed via $f(x)=\psi(x)^2$,
FIM adopts the simpler aspect \cite{frieden3} \be \label{eq.1-2} I
\,=\,\int ~dx ~f(x)\left\{\frac{\partial ~ }{\partial
x}~\ln{[f(x)]}\right\}^2=4\,\int
dx\,\left[\psi'(x)\right]^2;\,\,\,\,\,(d\psi/dx= \psi').\ee Note
that for the uniform distribution $f(x)=constant$ one has $I=0$.
Focus attention now a system that is specified by a set of $M$
physical parameters $\mu_k$. We can write $\mu_k = \langle
A_{k}\rangle$ with  $A_{k}= A_{k}(x).$ The set of $\mu_{k}$-values
is to be regarded as our prior knowledge. It represents available
empirical information. Let the pertinent probability distribution
function (PDF) be $f(x)$. Then, \be \label{eq.1-3} \langle
A_{k}\rangle\,=\,\int ~dx ~A_{k}(x) ~f(x), \hspace{0.5cm}
k=1,\dots ,M. \ee In this context it can be  shown (see for
example
  \cite{pla7,reginatto}) that the {\it physically relevant}
  PDF $f(x)$  minimizes the FIM (\ref{eq.1-2}) subject to
  the prior conditions and the normalization condition.

\nd In the celebrated MaxEnt approach of Jaynes \cite{katz} one
{\it maximizes} the entropy, that behaves information-wise in
opposite fashion to that of Fisher's measure \cite{frieden3}.
Normalization entails $\int dx  f(x) = 1,$  and, consequently, our
Fisher-based extremization problem adopts  the appearance \be
\label{eq.1-4}\delta \left( I - \alpha \int ~dx ~f(x) -
\sum_{k=1}^M~\lambda_k\int ~dx ~A_{k}(x)~f(x)\right) = ~0 \ee
where we have introduced the $(M+1)$ Lagrange multipliers
$\lambda_k$ ($\lambda_0=\alpha$). In Ref. \cite{pla7} on can find
the details of how to go  from (\ref{eq.1-4}) to a Schr\"odinger's
equation (SE) that yields the desired PDF in terms of the
amplitude $\psi(x)$ referred to above [i.e., before Eq.
(\ref{eq.1-2})]. This SE is of the form

 \be \label{eq.1-5} -~\frac{1}{2}~\frac{\partial^2 ~}{\partial x^2} \psi~-~\sum_{k=1}^{M}~\frac{\lambda_{k}}{8}~ A_{k}\,\psi ~=
 ~ \frac{\alpha}{8}~ \psi, \ee
and can be formally interpreted as the (real) Schr\"odinger
equation for a particle of unit mass ($\hbar=1$) moving in the
effective, ``information-related pseudo-potential"  \cite{pla7}
\be \label{eq.1-6} U~=~U(x)
=~-\frac{1}{8}~\sum_{k=1}^{M}\,\lambda_{k}~ A_{k}(x), \ee in which
the normalization-Lagrange multiplier ($\alpha /8$) plays the role
of an energy eigenvalue. The  $\lambda_k$ are fixed, of course, by
recourse to the available prior information. Note that $\psi(x)$
is always real in the case of one-dimensional scenarios, or for
the ground state of a real potential in N dimensions
\cite{richard}. In terms of the amplitudes $\psi(x)$ we have

\ben \label{eq.1-7}  I & = & \,\int dx ~f \left(\frac{\partial
\ln{f} }{\partial x}\right)^2\,= \, \int dx ~ \psi_n^2 ~
\left(\frac{\partial \ln{\psi_n^2} }{\partial x} \right)^2\,=\, 4
\int d x ~ \left(\frac{\partial \psi_n }{\partial x} \right)^2
=\n \\
& =&\, -~4 \int \psi_n \frac{\partial^2 ~}{\partial x^2} \psi_n~dx
=\, \int ~ \psi_n \left(\alpha + \sum_{k=1}^M~\lambda_k~A_k\right)
\psi_n~dx,\n \een

\nd i.e.,\ben \label{eq.1-12} I=\,\alpha
 + \sum_{k=1}^M~\lambda_k\left\langle A_k\right\rangle. \een
 a  form that we will employ in our developments below.
 Some useful
results of  Refs. \cite{Univ,nuestro2} will be needed below. An
essential ingredient in the present considerations is the {\it
virial theorem} \cite{virial} that, of course, applies in this
Schr\"odinger-scenario \cite{greiner}. It states that
 \ben \label{virial-4} \left\langle - ~ \frac{\partial^2
~}{\partial x^2}\right\rangle = \left\langle {x} ~ \frac{\partial
~}{\partial x} U({x})\right\rangle. \een  The potential function
$U(x)$ belongs to $\mathcal{L}_2$
 and thus admit of a series expansion in
$x,\,x^2\,x^3,\,$etc. \cite{greiner}. The $A_k(x)$ themselves
belong to $\mathcal{L}_2$ as well and can be series-expanded in
similar fashion. This enables us to base our future considerations
on the assumption that the a priori knowledge refers to moments
$x^k$ of the independent variable, i.e.,

\be \langle A_k \rangle~=~ \langle x^k \rangle ~,  \ee and that
one possesses information on  $M$ moment-mean values
 $\langle x^k \rangle$. Our ``information" potential $U$
  then reads

\be \label{virial-5} U(x)=  -~ \frac{1}{8} \sum_k \,\lambda_k\,
x^k. \ee  {\it We will assume that the first $M$ terms of the
above series yield a satisfactory
 representation of} $U(x)$. Consequently, the following
 identification is made

 \be \label{identif}  {\rm Lagrange\,\, multipliers}  \Leftrightarrow   {\rm U(x)'s\,\, series-expansion's\,\, coefficients}.  \ee
 Thus, Eq. (\ref{virial-4}) allows one to immediately
obtain

\ben \label{virial-6}
 \left\langle  \frac{\partial^2 ~}{\partial x^2}\right\rangle \,  =\, ~
\frac{1}{8}~\sum_{k=1}^{M}\, k \,\lambda_{k}~\left\langle
A_{k}\right\rangle;\hspace{1.2cm}(A_k=x^k),   \een and thus, via
 (\ref{virial-6}) and the above mentioned relation
 $I=-~4 \left\langle \frac{\partial^2 ~}{\partial x^2}
\right\rangle$, a useful, virial-related expression for Fisher's
information measure can be arrived at

 \ben \label{virial-7} I\,
=\, -~ ~\sum_{k=1}^{M}\, \frac{k}{2} \,\lambda_{k}~\langle
x^{k}\rangle, \een which is an explicit function of the M physical
parameters $\langle x^{k}\rangle$ and their respective Lagrange
multipliers (also, $U(x)$'s series-expansion's coefficients)
$\lambda_{k}$. Eq. (\ref{virial-7}) encodes the information
provided by the virial theorem \cite{Univ,nuestro2}. Note that if
we define $M-$dimensional vectors ${\bf X}$ of components
$X_k=<x^k>$ and ${\bf G}$ of components $G_k=k\,\lambda_k/2$ we
can cast $I$ in the scalar-product fashion

\be \label{1new}  I=- {\bf X}\cdot{\bf G}.\ee

\section{3. The Legendre structure}

\nd The connection between our variational solutions $f$  and
thermodynamics
  was established in Refs. \cite{pla7} and \cite{flego} in the guise of
  reciprocity relations that  express  the
Legendre-transform structure of thermodynamics. They constitute
its essential formal ingredient \cite{deslog} and  were re-derived
\`a la Fisher in \cite{pla7}  by recasting  (\ref{eq.1-12}) in a
fashion that emphasizes the role of the relevant independent
variables

\ben \label{eq.1-13a} I(\left\langle
A_1\right\rangle,\ldots,\left\langle A_M\right\rangle) \,=\,\alpha
 + \sum_{k=1}^M~\lambda_k\left\langle
A_k\right\rangle. \een The Legendre transform changes the identity
of our relevant variables. As for  $I$ we have

\be \label{eq.1-13b}\alpha= I(\left\langle
A_1\right\rangle,\ldots,\left\langle A_M\right\rangle) -
\sum_{k=1}^M~\lambda_k\left\langle A_k\right\rangle =
\alpha(\lambda_1,\ldots,\lambda_M), \ee so that we encounter the
three reciprocity relations  proved in \cite{pla7}

\be \label{RR-1} \frac{\partial \alpha}{\partial \lambda_{i}}= -
\langle A_i\rangle ~; \hspace{1.cm}  \frac{\partial I }{\partial
\left\langle A_k \right\rangle}\,=\,\lambda_k  ~ ;\hspace{1.cm}
\frac{\partial I}{\partial \lambda_{i}}=\sum_{k}^{M} \lambda_{k}
 \frac{\partial \langle A_{k}\rangle}{\partial \lambda_{i}},\ee
the last one being a generalized Fisher-Euler theorem. From
(\ref{eq.1-13b}) and (\ref{RR-1}), one can obtain an infinite set
of relations linking $I$ and $\alpha$ by taking derivatives of
(\ref{eq.1-13b}) with respect to $\lambda_k$ or $\langle
A_k\rangle$. For example, the relation between the second
derivatives is given by
 \ben \label{RR-2} \sum_{k=1}^M{~\left(\frac{\partial^2
I}{\partial \langle A_i\rangle\partial \langle A_k\rangle
}\right)\left(\frac{\partial^2 {\alpha}}{\partial \lambda_k
\partial \lambda_j}\right)}~=~-~ \sum_{k=1}^M{~\left(\frac{\partial
\lambda_k }{\partial \langle A_i\rangle} \right)
\left(\frac{\partial \langle A_j\rangle }{\partial \lambda_k
}\right)}~=~-~ \delta_{ij}. \een where $\delta_{ij}$ is the unit
matrix.  
FIM expresses a relation between the independent variables or
control variables (the prior information) and a dependent value
$I$. Such information is encoded into the functional form of
$I=I(\langle A_1 \rangle, ... , \langle A_M \rangle  )$. For later
convenience, we will also denote such a relation or encoding as
$\{I,\langle A_k \rangle \}$. We see that the Legendre transform
FIM-structure involves eigenvalues of the information-Hamiltonian
 which display the information encoded in
 $I$ via Lagrange multipliers,
${\alpha}={\alpha}(\lambda_1, ... \lambda_M)~ :$ \ben \label{RR-3}
\{I,\langle A_k \rangle \} \hspace{0.6cm} \longleftrightarrow
\hspace{0.6cm} \{{\alpha}, \lambda_k \}. \n \een
\section{4. Main results}

\nd We start here with our present developments. Substituting
(\ref{virial-7}) into (\ref{eq.1-12}) and solving for $\alpha$, we
obtain \ben \label{gov-a} \alpha \,  =\,
-~\sum_{k=1}^{M}\,\left(1+\frac{k}{2}\right) ~\lambda_k ~\langle
x^k \rangle.  \een Since $\langle x^k \rangle$ is given by
(\ref{RR-1}) as $[-
\partial \alpha/\partial \lambda_k ],$ (\ref{gov-a}) take the form
\ben \label{gov-a1} \alpha \,  =\,
~\sum_{k=1}^{M}\,\left(1+\frac{k}{2}\right) ~\lambda_k ~
\frac{\partial \alpha }{\partial \lambda_k} \,.\een  \nd
 Eq.
(\ref{gov-a1}) constitutes  an important result, since we have now
at our disposal  a {\it linear, partial differential equation
(PDE)} for $\alpha$, whose variables are $U(x)$'s
series-expansion's coefficients 
 The equation's
origins are two information sources, namely, i) the Legendre
structure and ii) the virial theorem. Dealing with this new
equation might allow us to find $\alpha$ in terms of the
$\lambda_k$ {\it without passing before through a Schr\"odinger
equation}, a commendable achievement. See below, however, the
pertinent caveats. \vskip 3mm \nd For convenience we now recast
our key relations using dimensionless magnitudes \ben
\label{gov-2a}
\mathcal{A}~=~\frac{\alpha}{[\alpha]}~=~\frac{\alpha}{[x]^{-2}}~
\hspace{0.2cm},\hspace{1.2cm} \Lambda_k ~=~
\frac{\lambda_k}{[\lambda_k]} ~=~ \frac{\lambda_k}{[x]^{-(2+k)}
}~,\hspace{0.5cm}\een
 where $[\alpha]$ and $[\lambda_k]$ denote the dimensions of
$\alpha$  and $\lambda_k$, respectively. Thus, the differential
equation that governs the energy-behavior, i.e., (\ref{gov-a1}),
can be translated into \ben  \label{gov-3a} \mathcal{A} \,  =\,
\sum_{k=1}^{M}\,\left(1+\frac{k}{2}\right) ~{\Lambda}_{k} ~
\frac{\partial \mathcal{A}}{\partial \Lambda_k}\,, \een and is
easy to obtain a solution as follows. One sets
  \ben \label{gov-4a}
\mathcal{A}=~\sum_{k=1}^M~\mathcal{A}_k ~ =
  ~\sum_{k=1}^M~ \exp{\left[~ h(\Lambda_k )\right]},\een
   and substitution of (\ref{gov-4a}) into (\ref{gov-3a}) leads to

\ben \label{gov-5a} \mathcal{A} \,  =\,
\sum_{k=1}^{M}\,\left(1+\frac{k}{2}\right) ~\Lambda_{k} ~
h'(\Lambda_k )~\mathcal{A}_k\,.\een  The above relation entails
 \ben \label{gov-6a} ~
h'(\Lambda_k )= ~\frac{2}{(2+k)}\frac{1}{~\Lambda_k }
\hspace{1.cm}\longrightarrow \hspace{1.cm} ~h(\Lambda_k ) =
~\frac{2}{2+k}~\ln{\left| \Lambda_{k}\right|}+d_k \, , \een where
$d_k$ is an integration constant.  Finally, inserting
(\ref{gov-6a}) into (\ref{gov-4a}) we arrive at

 \ben
\label{gov-7a}
\mathcal{A}=\sum_{k=1}^{M}~D_k~\exp{\left(~\frac{2}{2+k}~
\ln{\left|\Lambda_k \right|} \right)}~ , \hspace{1.2cm}D_k =~
e^{d_k}~>~0~,\een  which can be recast as \ben \label{gov-8a}
\mathcal{A}(\Lambda_1, ... , \Lambda_M \rangle ) =~
\sum_{k=1}^{M}~D_k~ ~{ \left|
 \Lambda_{k} \right|^{{2}/{(2+k)}}}~ , \een or, in
function of the original input-quantities (\ref{gov-2a})

\ben \label{gov-9a} {\alpha}(\lambda_1, ... , \lambda_M ) = ~
\sum_{k=1}^{M}{\alpha_k}(\lambda_k)~= ~ \sum_{k=1}^{M}~D_k~ ~{
\left| \lambda_k \right|^{{2}/{(2+k)}}}~,\een implying what seems
to be a universal prescription, a linear PDE, that energy
eigenvalues must necessarily comply with.  This constitutes one of
the main present results. Of course, our solution poses a
necessary but not (yet) sufficient condition for $\alpha$ to be an
energy-eigenvalue.

\nd All first order, linear PDEs possess a solution that depends
 on an arbitrary function, called {\it the general solution} of the PDE.
In many physical situations this solution if {\it less important}
than other solutions called {\it complete ones}
\cite{pde,pde-1,pde-3}. Such complete solutions are particular PDE
solutions containing as many arbitrary constants as intervening
independent variables. As an example we may cite the integration
of the classical equations of motion via a methodology involving
Hamilton-Jacobi equations, for which a complete integral is
required \cite{pde,pde-1,pde-3}.
  We will delve into this question again in Section 8 and obtain the
general solution of our PDE. In Sec. 9 we will discuss its
uniqueness via analysis of the associated  Cauchy problem.

\section{5. Main properties of  $\alpha$}

\nd Some important properties deserve special mention.
\begin{itemize}
    \item{\bf $\alpha$-domain}

\nd    Obviously, it is
\[{\it Dom}[\alpha]=\left\{(\lambda_1,\cdots, \lambda_M ) / \lambda_k ~\in ~\Re \right\}=\Re^M\]

    \item{\bf $\alpha$-monotonicity}

Differentiating (\ref{gov-9a}) we obtain \ben \label{prop-2a}
 \frac{\partial \alpha}{\partial \lambda_k} ~=~
\frac{2}{(2+k)~\lambda_k} ~ \alpha_k ~=~ \frac{2}{(2+k)~\lambda_k}
~ D_k~ ~ \left| \lambda_k \right|^{2/(2+k)} \een Therefore, if $
\lambda_k ~ < ~ 0 ~$ , $\alpha$ is a monotonically decreasing
function in the $ \lambda_k$-direction.  Also, for $ \lambda_k ~ <
~ 0 ~$, from the reciprocity relations (\ref{RR-1}) we have, \ben
\label{prop-2a-RR}
 \langle x^k \rangle~=~-~\frac{\partial \alpha}{\partial \lambda_k} ~=~
\frac{2}{(2+k)}~ D_k~ ~ \left| \lambda_k \right|^{-~k/(2+k)} ~
>~0. \een

\item{\bf $\alpha$-convexity}

\nd This is a necessary property, since the $\tilde{\alpha}= -
~\alpha$ is the  Legendre transform of FIM.

 By differentiation of the expression (\ref{prop-2a}) one obtains \ben
 \label{prop-3a} \frac{\partial^2 \alpha }{\partial
\lambda_n \partial \lambda_k} \,=\, -~ \frac{2 k}{(2+k)^2}~ D_k~
\left| \lambda_k \right|^{~-~2(1+k)/(2+k)} ~\delta_{kn}~,\een from
which we can assert that $\alpha$ is concave and, obviously,
$\tilde{\alpha}= - \alpha$ is  a convex function. It is then
guaranteed that the inverse transform of $\partial_n \partial_k
{\alpha}$ exists.
\end{itemize}
We end this section by mentioning that an $I-$analog of  Eq.
(\ref{gov-9a}) exists, namely, \ben \label{gov-9} {I}(\langle
{x}^1 \rangle, ... , \langle {x}^M \rangle ) =~
\sum_{k=1}^{M}~I_k~ = ~ \sum_{k=1}^{M}~C_k~ ~{ \left| \langle
{x}^{k}\rangle \right|^{- {2}/{k}}}~,\een
 where $C_k$ are positive real number (integration constant). Eq. (\ref{gov-9}) constitutes the main result of Ref. \cite{Univ}.
 We are going to enumerate below some properties can be directly derived
from it, relevant for the present work.

\begin{itemize}
    \item{\bf FIM-domain}

\nd    Obviously, it is
     \[{\it Dom}[I]=\left\{(\langle {x}^1 \rangle, ... , \langle {x}^M
\rangle) / \langle {x}^k \rangle~\in ~\Re_o \right\}\]

    \item{\bf FIM-monotonicity}

Differentiating (\ref{gov-9}) one obtain \ben \label{prop-2}
\frac{\partial I}{\partial \langle x^k \rangle} ~=~-~
\frac{2}{k~\langle x^k \rangle}I_k~=-~ \frac{2}{k~\langle x^k
\rangle}~C_k~ ~{ \left| \langle {x}^{k}\rangle \right|^{-~ 2/k}}~,
\een Therefore, if $\langle x^k \rangle ~ > ~ 0 ~$ , $I$ is a
monotonically decreasing function in the $ \langle x^k
\rangle$-direction.  Also, for $\langle x^k \rangle ~ > ~ 0 ~$,
from the reciprocity relations (\ref{RR-1}) one have,
 \ben
\label{prop-2-RR} \lambda_k ~=~ \frac{\partial I}{\partial \langle
x^k \rangle} ~=~-~ \frac{2}{k}~C_k~ ~{ \langle {x}^{k}\rangle^{-~
(2+k)/k}}~<~0~. \een

    \item{\bf FIM-convexity}

\nd  By differentiation of the expression (\ref{prop-2}) one
obtains
 \ben  \label{prop-3} \frac{\partial^2 I }{\partial
\langle x^n \rangle\partial \langle x^k \rangle}
 \,=\, \left( \frac{2+k}{2} \right)  \frac{4}{k^2} ~ C_k~\left|\langle x^k \rangle \right|^{-~2(1+k)/k}~\delta_{kn},
\een from which we can assert that the Fisher measure is  a convex
function. It is then guaranteed that the inverse of $\partial_k
\partial_j \bar{\alpha}$ exists.

\end{itemize}


\section{6. The mathematical structure of the Legendre transform}

\nd In order to better understand the formalism developed in the
preceding Section  we scrutinize now in some detail  the
mathematical structure associated to the Legendre transform (see
(\ref{eq.1-13b}), (\ref{RR-1}) and (\ref{RR-2})). This  leads to a
relation between the integration constants $C_k$ and $D_k$
pertaining to the $I$ and $\alpha$ expressions given by
(\ref{gov-9}) and (\ref{gov-9a}). We are going to study this
relation in both scenarios, $\left\{\alpha, \lambda_k \right\}$
and $\left\{I, \langle {x}^{k}\rangle \right\}$. Remember that our
Lagrange multipliers are simultaneously $U(x)$'s
series-expansion's coefficients. \vskip 3mm

\nd {\bf In a  $\left\{I, \langle {x}^{k}\rangle \right\}$ -
scenario},
 the $\lambda_k$ are  functions dependent on the $\langle
 {x}^{k}\rangle$-values. Taking into account  (\ref{prop-2-RR}), the energy
(\ref{gov-9a}) and the potential, expressed in function of the
independent  $\langle {x}^{k}\rangle$-values, take the form

\ben \label{IX-1}
\alpha~=~
\sum_{k=1}^{M} D_k{ \left| \lambda_k \right|^{{2}/{(2+k)}}}~=
\sum_{k=1}^M D_k\left(\frac{2}{k}~C_k\right)^{2/(2+k)}  \left| \langle x^k\rangle \right|^{- 2/k}~, \een

\ben \label{IX-2} \sum_{k=1}^M \lambda_k \langle x^k \rangle~
=~-~\sum_{k=1}^M \frac{2}{k}~C_k \left| \langle {x}^{k}\rangle
\right|^{- 2/k}~. \een  Substituting (\ref{gov-9}), (\ref{IX-1})
and (\ref{IX-2}) into (\ref{eq.1-12}) we have

\ben \label{IX-3} \sum_{k=1}^M C_k~\left| \langle x^k\rangle
\right|^{- 2/k} ~=~\sum_{k=1}^{M} D_k \left(~\frac{2}{k} C_k
\right)^{2/(2+k)}   \left| \langle x^k\rangle \right|^{- 2/k} -
\sum_{k=1}^{M} \frac{2}{k} C_k  \left| \langle x^k\rangle
\right|^{- 2/k}\, ,\n \een \nd which can be recast as \ben
\label{IX-4} \sum_{k=1}^{M}
\left\{~D_k~\left(~\frac{2}{k}~C_k~\right)^{2/(2+k)}-~
\frac{2+k}{k}~C_k~\right\}\left| \langle x^k\rangle \right|^{-
2/k}~=~0~.  \een \nd The above equation is automatically fulfilled
if we impose that \ben \label{IX-5}
~D_k~\left(~\frac{2}{k}~C_k~\right)^{2/(2+k)} ~=~
~\frac{2+k}{k}~C_k~~,\n \een which leads to \ben \label{IX-6}
~D_k~C_k^{-k/(2+k)}~=~\frac{(2+k)}{2}
~\left(~\frac{k}{2}\right)^{-k/(2+k)}~. \een

\nd We can verify that the above relation between $C_k$ and $D_k$
preserves the symmetric representation of the second derivatives
(\ref{RR-2}). Using  (\ref{prop-2-RR}) we can express
(\ref{prop-3a}) as a function of the $\langle x^k \rangle$,

 \ben
\frac{\partial^2 \alpha }{\partial \lambda_k \partial \lambda_n}  &=&
- \left(\frac{2}{k}\right)^{-1}
\left(\frac{2}{2+k}\right)^2~ D_k~\left|\lambda_k\right|^{-2(1+k)/(2+k)} ~\delta_{kn}
=\n \\
\label{IX-7} &=&-\left(\frac{2}{k}\right)^{-(4+3k)/(2+k)}
\left(\frac{2}{2+k}\right)^2 D_k~C_k^{-2(1+k)/(2+k)} \left|
\langle {x}^{k}\rangle \right|^{2(1+k)/k}
~\delta_{kn}~.\hspace{1.cm}\een  The sum over $k$ of the product
of (\ref{IX-7}) and (\ref{prop-3}) leads to

\ben
 \label{IX-8} \sum_{k=1}^{M}{
\frac{\partial^2 I }{\partial \langle x^l \rangle \partial \langle
x^k \rangle} \frac{\partial^2 \alpha }{\partial \lambda_k \partial
\lambda_n}}\,=\,-~ \sum_{k=1}^{M}~{ \left(
\frac{2}{k}\right)^{-k/(2+k)} \frac{2}{2+k}~  C_k^{-k/(2+k)}~D_k
~\delta_{kn} \delta_{lk} } , \een  which, using (\ref{IX-6})
reduces to

\ben
 \label{IX-9} \sum_{k=1}^{M}{
\frac{\partial^2 I }{\partial \langle x^l \rangle \partial \langle
x^k \rangle} \frac{\partial^2 \alpha }{\partial \lambda_k \partial
\lambda_n}}\,=\,-~ \sum_{k=1}^{M}{ ~\delta_{kn}~ \delta_{lk} }
=~-~ \delta_{ln}, \een \nd as  expected from  (\ref{RR-2}).

\vskip 3mm

   \nd {\bf In the $\left\{\alpha, \lambda_k \right\}$ scenario},
   the $\langle {x}^{k}\rangle$ are  functions that depend  on the
   $\lambda_k$-values.    \nd Taking into account i) (\ref{prop-2a-RR}),
    ii) the FIM-relation (\ref{gov-9}),  and iii) the information-potential, expressed as a function of the independent
      $\lambda_k$-values, FIM adopts the appearance

\ben \label{AL-1} I
~ = ~ \sum_{k=1}^{M}~C_k~ ~{ \left| \langle {x}^{k}\rangle \right|^{- {2}/{k}}}~~=~
\sum_{k=1}^M ~~\left(~\frac{2}{2+k}\right)^{-2/k} ~C_k~D_k^{~-2/k} ~ \left| \lambda_k \right|^{2/{(2+k)}}~, \een

\ben \label{AL-2} \sum_{k=1}^M {~\lambda_k \langle x^k \rangle}~
=~-~\sum_{k=1}^M \frac{2}{(2+k)}~D_k~ ~{ \left| \lambda_k
\right|^{{2}/{(2+k)}}}. \een \nd Substituting (\ref{gov-9a}),
(\ref{AL-1}), and (\ref{AL-2}) into (\ref{eq.1-12}) we have

\ben \label{AL-3} \sum_{k=1}^M C_k
\left(~\frac{2~D_k}{2+k}\right)^{-2/k}\left| \lambda_k
\right|^{2/{(2+k)}}=\sum_{k=1}^{M} D_k \left| \lambda_k
\right|^{2/(2+k)} - \sum_{k=1}^{M}\, \frac{2}{(2+k)} D_k \left|
\lambda_k \right|^{2/(2+k)}~,\n \een \nd which can be recast as
\ben \label{AL-4} \sum_{k=1}^{M}
\left\{~C_k~\left(~\frac{2~D_k}{2+k}\right)^{-2/k}~-~
\frac{k}{(2+k)}~D_k~\right\}\left| \lambda_k
\right|^{2/(2+k)}~=~0~.\n \een

\nd The above equation is automatically fulfilled if we enforce
\ben \label{AL-5} ~C_k~\left(~\frac{2~D_k}{2+k}\right)^{-2/k}=
\frac{k}{(2+k)}~D_k~,\n \een \nd which leads to \ben \label{AL-6}
~C_k~D_k^{-(k+2)/k}~=~\frac{k}{2}
~\left(~\frac{2+k}{2}\right)^{-(k+2)/k}~. \een

\nd We can verify that the above  relation between $C_k$'s and
$D_k$'s preserves the symmetric representation of the second
derivatives (\ref{RR-2}). Using  (\ref{prop-2a-RR}) we can express
(\ref{prop-3}) as a  function of the $\lambda_k$ \ben
\frac{\partial^2 I }{\partial \langle x^l \rangle \partial \langle
x^k \rangle} &=&\frac{(2+k)}{2}~  \frac{4}{k^2}  ~C_k~
\left| \langle x^k \rangle\right|^{-2(1+k)/k}~ \delta_{lk}\n \\
&=&\frac{(2+k)}{2}  \frac{4}{k^2}  ~C_k~\left(\frac{2D_k}{2+k}\right)^{-2(1+k)/k}
 \left| \lambda_k \right|^{2(1+k)/(2+k)}~ \delta_{lk}\n \\
\label{AL-7}
  &=&\left(\frac{2}{k}\right)^2\left(\frac{2}{2+k}\right)^{-(2+3k)/k}  C_k~D_k^{-2(1+k)/k}
 \left| \lambda_k \right|^{2(1+k)/(2+k)}~ \delta_{lk}. \een
\nd The sum over $k$ of the product of (\ref{prop-3a}) and
(\ref{AL-7}) now gives

\ben
 \label{AL-8} \sum_{k=1}^{M}{
\frac{\partial^2 I }{\partial \langle x^l \rangle \partial \langle
x^k \rangle} \frac{\partial^2 \alpha }{\partial \lambda_k \partial
\lambda_n}}\,=\,-~ \sum_{k=1}^{M}~{  \frac{2}{k}
\left(\frac{2}{2+k}\right)^{-(2+k)/k}  C_k~D_k^{-(2+k)/k}
~\delta_{kn} \delta_{lk} } , \een which, using (\ref{AL-6}),
reduces to \ben
 \label{AL-9} \sum_{k=1}^{M}{
\frac{\partial^2 I }{\partial \langle x^l \rangle \partial \langle
x^k \rangle} \frac{\partial^2 \alpha }{\partial \lambda_k \partial
\lambda_n}}\,=\,-~ \sum_{k=1}^{M}{ ~\delta_{kn}~ \delta_{lk} }
=~-~ \delta_{ln}, \een as we expect from (\ref{RR-2}).
 \nd From Eqs. (\ref{IX-6}) or (\ref{AL-6}) we can write

\ben \label{IX-AL-1} ~C_k~=~\frac{k}{2}~\bar{C}_k~,\hspace{2.cm}
D_k=~\frac{k+2}{2}~\bar{D}_k~ \een \nd with \ben \label{IX-AL-2}
~\bar{D}_k^{~(2+k)}~=~\bar{C}_k^{~k}~. \een

\nd Now expressions (\ref{gov-9}) and (\ref{gov-9a})  take the
form,

\ben \label{IX-AL-3} {I} &=&  \sum_{k=1}^{M} ~\frac{k}{2} ~\bar{C}_k
 ~{ \left|\langle {x}^{k}\rangle \right|^{-2/k}}~,\\
 &&\n \\
  \label{IX-AL-4}
\alpha &=& \sum_{k=1}^{M}~\frac{k+2}{2}~\bar{D}_k~
~\left|\lambda_k \right|^{2/(2+k)}~,\een \nd and the reciprocity
relations (\ref{prop-2-RR}) and (\ref{prop-2a-RR}) are given by

 \ben
\label{IX-AL-5} \lambda_k &=& \frac{\partial I}{\partial \langle
x^k \rangle} ~=~-\bar{C}_k~{ \langle {x}^{k}\rangle^{-~ (2+k)/k}}~,\\
 &&\n \\
\label{IX-AL-6}
 \langle x^k \rangle&=&-~\frac{\partial \alpha}{\partial \lambda_k} ~=
 ~~\bar{D}_k~ \left| \lambda_k \right|^{-~k/(2+k)}. \een

\nd Also, we can write
\ben \label{IX-AL-7} ~\bar{D}_k^{~(2+k)}~=~\bar{C}_k^{~k}~\equiv F_k^{~2} \een
\nd then, the expressions (\ref{gov-9}) and (\ref{gov-9a}),  take the form,
\ben \label{IX-AL-8} {I} &=&  \sum_{k=1}^{M}~\frac{k}{2}
 ~{ \left[\frac{F_k}{\langle {x}^{k}\rangle} \right]^{2/k}}~,\\
 &&\n \\
  \label{IX-AL-9}
{\alpha} &=& \sum_{k=1}^{M}~\frac{k+2}{2}~
{\left[F_k~\left|\lambda_k \right|\right]^{{2}/{(2+k)}}}~.\een \nd
and the reciprocity relations (\ref{prop-2-RR}) and
(\ref{prop-2a-RR}) can be summarized as \ben \label{IX-AL-10}
F_k^{~2}~=~ \left|\lambda_k \right|^{k}~\langle
{x}^{k}\rangle^{(2+k)}.\een

 \nd As was conjectured in \cite{Univ}, the reference-quantities $F_k$
 should contain important information concerning the referential system with respect the
which prior conditions are experimentally determined. Following
ideas advanced in  \cite{Univ} we will  look for the ``point" at
which the potential function achieves a minimum.

\section{7. Appropriate referential system}

\nd { \bf Minimum of the information potential}

\nd It is convenient to incorporate at the outset, within the $I-$
 and  $\alpha-$forms, information concerning the minimum of the
information potential. Assume that this potential \ben
U(x)=-~\frac{1}{8}\sum_{x=1}^M \lambda_k x^k, \n \een achieves its
absolute minimum at the ``critical point" $x=\xi$ \ben
U^{~'}(\xi)~=~0~, \hspace{2.cm} U_{min}~=~U(\xi).\een \nd Thus,
effecting the translational transform $u=x-\xi $ leads us to \ben
\label{sys-1} I=-~ \sum_{k=1}^{M}~\frac{k}{2}~\lambda_k~ \langle
{x}^{k}\rangle = ~-~ \sum_{k=1}^{M}~\frac{k}{2}~\lambda_k^*~
\langle {u}^{k}\rangle', \een with (see the Appendix) \ben
 \label{sys-2} \lambda_k^*~= ~-~\frac{8}{k!}~U^{(k)}(\xi)~, \hspace{1.2cm}
  \langle {u}^{k}\rangle' \,= \,\langle (x-\xi)^{k}\rangle~  \een
where $U^{(k)}(\xi)$ is the $k^{th}$ derivative of U(x) evaluated
at $x=\xi$ and $\langle ~ \rangle'$ indicates that the relevant
moment (expectation)  is computed with translation-transformed
eigenfunctions.

\begin{itemize}
    \item The corresponding FIM-explicit functional expression is built up with
the $N-$non-vanishing momenta ($N < M$) ($\langle u^k \rangle'
\neq 0$) and is given by
\ben \label{sys-3} {I} = ~
\sum_{k=2}^{N}~\frac{k}{2}~\bar{C}_k~{ \left| \langle {u}^{k}\rangle' \right|^{
-2/k}}~= ~ \sum_{k=2}^{N}~\frac{k}{2}~\bar{C}_k~{ \left| \langle (x-\xi)^{k}\rangle
\right|^{ -2/k}}~,\een \nd where we kept in mind that
$\lambda_1^*= - 8 U'(\xi)~=~0.$  A glance at the above
FIM-expression suggests that we  re-arrange things in the fashion
\ben \label{CR-1} {I} = ~\bar{C}_2~
\left|\langle (x-\xi)^{2}\rangle\right|^{-1}+ \sum_{k=3}^{N}~\frac{k}{2}~\bar{C}_k~ \left|\langle (x-\xi)^{k}\rangle
\right|^{- 2/k}~.\een Taking now into account
that

\ben \label{CR-2}
\left\{ \begin{array}{l}
\langle~ x-\xi~\rangle =0\\
\langle (x-\xi)^{2}\rangle =\langle x^{2}\rangle-2 \xi \langle x\rangle+\xi^2\\
\end{array} \right. \hspace{0.4cm}\longrightarrow\hspace{0.4cm}
 \left\{\begin{array}{l}
\langle~ x~\rangle ~=~\xi\\
\langle (x-\xi)^{2}\rangle =\langle x^{2}\rangle- \langle x\rangle^2=\sigma^2
\end{array} \right. \hspace{0.7cm}
~\een  we get \ben \label{CR-3} {I} =   ~\bar{C}_2~{\sigma^2}~+
\sum_{k=3}^{N}~\frac{k}{2}~\bar{C}_k~ \left|\langle
(x-\xi)^{k}\rangle \right|^{- 2/k}~,\een from which we obtain \ben
\label{CR-4} {I}~\sigma^2 = ~\bar{C}_2~+ ~\sigma^2
\sum_{k=3}^{N}~\frac{k}{2}~\bar{C}_k~ \left|\langle
(x-\xi)^{k}\rangle \right|^{- 2/k} ~\geq ~1~.\een Therefore, if no
moment $k \ge3$ is a priori known, in forcing $I$ to preserve the
well-known Cramer-Rao $I-$bound \cite{frieden3} $
I~{\sigma^2}~\geq~ 1$, we need that
\[\bar{C}_2=1 \hspace{1.5cm}\longrightarrow  \hspace{1.5cm}     \bar{C}_2=\bar{D}_2=F_2=1. \]

\item The
corresponding $\alpha$-explicit functional expression is
constructed with the $N-$non-vanishing momenta ($N < M$) ($\langle
u^k \rangle' \neq 0$) and is given by \ben \label{alph-3} \alpha
~= ~8~U(\xi) +\sum_{k=2}^{N}~\frac{k+2}{2}~\bar{D}_k~{ \left|
\lambda_k^*\right|^{ 2/(k+2)}}~.\een \nd For the harmonic
oscillator it is well known that \cite{nuestro2,Univ}

\ben \label{ES-2} U(x)=-~ \frac{1}{8}\lambda_2~x^2 ~,
\hspace{1.5cm}\lambda_2 \,= ~ - ~ 4\omega^2. \een \nd The minimum
of the potential function is obtained at the origin $\xi = 0$,
\[V'(\xi)=-4 ~\lambda_2~\xi=0\hspace{1.cm}\longrightarrow\hspace{1.cm}
\xi=0.\] \nd Thus, using the $\alpha$-expression (\ref{IX-AL-4})
con $\bar{D}_2 = 1$, we have \ben \label{ES-3} \alpha ~=~
~2~{\left|\lambda_2 \right|^{1/2}}~=~4w~.\een \nd as we should
expect since  ($\alpha /8$) plays the role of an energy eigenvalue
[Cf. Eq. (\ref{eq.1-4})] and we took Planck's constant equal to
unity.

\end{itemize}


\section{8. General Solution of the energy-equation}

\nd Our $\alpha-$equation is a  first order linear nonhomogeneous
differential equations. We are following \cite{pde,pde-1,pde-3} in
looking for the general solution. For a first-order PDE, the {\it
method of characteristics}
 allows one to encounter useful curves (called characteristic curves or just
characteristics) along which the PDE becomes an ordinary
differential equation (ODE). Once the ODE is found, it can be
solved along the characteristic curves and transformed into a
solution for the original PDE.

\nd We are dealing with  a first order linear nonhomogeneous
equation with $M$ independent variables of the form (\ref{gov-2a})\ben
{\label{pde-s1}} \sum_{k=1}^M{\left(\frac{2+k}{2}\right)\Lambda_k}
\frac{\partial{\mathcal{A}}}{\partial \Lambda_k} ={\mathcal{A}}~,
\hspace{1.5cm}
\mathcal{A}=~\mathcal{A}\left(\Lambda_1,\cdots,\Lambda_M\right),
\een

\nd whose characteristic system
\ben {\label{pde-s2}} \frac{d\Lambda_i}{((2+i)/2)\Lambda_i
}=\frac{d\Lambda_j}{((2+j)/2)\Lambda_j}
=\frac{d\mathcal{A}}{\mathcal{A}}~,\hspace{1.cm}
i,j=1,\cdots,M,\een leads (for $\Lambda_1\neq 0$) to \ben
{\label{pde-s3}} \frac{d\Lambda_1}{(3/2)\Lambda_1
}=\frac{d\Lambda_k}{((2+k)/2)\Lambda_k}
\hspace{0.5cm}\longrightarrow\hspace{0.9cm}
\frac{2}{3}~\ln{\left|\Lambda_1\right|}+{c_1}&=&\frac{2}{2+k}~\ln{\left|\Lambda_k\right|}+{c_k}\n\\
\ln{\left[e^{c_1}~\left|\Lambda_1\right|^{2/3}\right]}&=&~\ln{~\left[e^{c_k}~\left|\Lambda_k\right|^{2/(2+k)}\right]}\n\\
e^{c_1}~\left|\Lambda_1\right|^{2/3}&=&~e^{c_k}~\left|\Lambda_k\right|^{2/(2+k)}\n\\
&\downarrow & \n \\
b_{k-1}\equiv
e^{c_1-c_k}&=&~\frac{\left|\Lambda_k\right|^{2/(2+k)}}
{\left|\Lambda_1\right|^{2/3}~}\\
\n\\
{\label{pde-s4}} \frac{d\Lambda_1}{(3/2)\Lambda_1
}=\frac{d\mathcal{A}}{\mathcal{A}}
\hspace{1.5cm}\longrightarrow\hspace{0.9cm}
\frac{2}{3}~\ln{\left|\Lambda_1\right|}+{c_1}&=&~\ln{\left|\mathcal{A}\right|}+{c_{\mathcal{A}}}\n\\
\ln{\left[e^{c_1}~\left|\Lambda_1\right|^{2/3}\right]}&=&~\ln{~\left[e^{c_{\mathcal{A}}}~\left|{\mathcal{A}}\right|\right]}\n\\
e^{c_1}~\left|\Lambda_1\right|^{2/3}&=&~e^{c_{\mathcal{A}}}~\left|{\mathcal{A}}\right|\n\\
&\downarrow & \n \\
b_M\equiv e^{c_1-c_{\mathcal{A}}}&=&~
\frac{\left|\mathcal{A}\right|}{\left|\Lambda_1\right|^{2/3}~}.
\een \nd We have now constructed an integral basis for the
characteristic system (\ref{pde-s2}) \ben {\label{pde-s5}}
b_1=u_1(\Lambda_1,...,\Lambda_M,\mathcal{A})~,  ~.~.~.~,~
b_M=u_{M}(\Lambda_1,...,\Lambda_M,\mathcal{A})~, \een \nd and the
general solution of equation (\ref{pde-s1}) defined as

\ben {\label{pde-s6}} \Phi(u_1,u_2,~.~.~.~,u_{M})~=~0, \een is
given by

\ben {\label{pde-s7}}
\Phi\left(\frac{\left|\Lambda_2\right|^{1/2}}{\left|\Lambda_1\right|^{2/3}},
\cdots,\frac{\left|\Lambda_k\right|^{2/(2+k)}}{\left|\Lambda_1\right|^{2/3}~},\cdots,\frac{\left|\Lambda_M\right|^{2/(2+M)}}{\left|\Lambda_1\right|^{2/3}~},\frac{\left|
\mathcal{A} \right|}{~~\left|\Lambda_1\right|^{2/3}~}\right)~=~0,
\een \nd where $\Phi$ is an arbitrary function of the $M$
variables. Solving this equation for $\mathcal{A}$ yields a
solution of the  explicit form

\ben {\label{pde-s8}} \mathcal{A}
~=~{~~\left|\Lambda_1\right|^{2/3}~}~\Psi
\left(\frac{\left|\Lambda_2\right|^{1/2}}{\left|\Lambda_1\right|^{2/3}},
\cdots,\frac{\left|\Lambda_k\right|^{2/(2+k)}}{\left|\Lambda_1\right|^{2/3}~},
\cdots,\frac{\left|\Lambda_M\right|^{2/(2+M)}}{\left|\Lambda_1\right|^{2/3}~}\right),
\een \nd where $\Psi$ is an arbitrary function of ($M-1$)
variables.

\section{9. Cauchy problem and the existence and uniqueness of the solution to our PDE}

\nd One of the fundamental aspects so as to have a useful  PDE  for modeling physical  systems
 revolves around the existence and  uniqueness of the solutions to the Cauchy problem  
 Here we  show that such  requirements are satisfied by our pertinent solutions.
\nd We start by casting (\ref{pde-s1}) in the normal form

\ben {\label{pde-s14}}
 \frac{\partial{\mathcal{A}}}{\partial \Lambda_1} =F\left(\Lambda_1,\cdots,\Lambda_M,{\mathcal{A}}, \frac{\partial{\mathcal{A}}}{\partial \Lambda_2},\cdots, \frac{\partial{\mathcal{A}}}{\partial \Lambda_M}\right)
\een where \ben
{\label{pde-s15}}F\left(\Lambda_1,\cdots,\Lambda_M,{\mathcal{A}},
\frac{\partial{\mathcal{A}}}{\partial \Lambda_2},\cdots,
\frac{\partial{\mathcal{A}}}{\partial
\Lambda_M}\right)=\frac{2}{3\Lambda_1}\left[\mathcal{A}-
\sum_{k=2}^M{ \frac{k+2}{2}\Lambda_k ~
\frac{\partial{\mathcal{A}}}{\partial \Lambda_k}} \right]~, \een
\nd and we see that $F$ is a real function of class $C^2$ in a
neighborhood of \ben {\label{pde-s16}} \Lambda_1=a~,
\hspace{0.3cm}\Lambda_k=\xi_{k-1}~, \hspace{0.3cm}
\mathcal{A}({\xi_1,...,\xi_{M-1}})=c~~, \hspace{0.3cm}
\left.\frac{\partial{\mathcal{A}}}{\partial
\Lambda_k}\right|_{\xi_1,...,\xi_{M-1}}=d_{k-1} ~~,
\hspace{0.5cm}k=2,\cdots,M \hspace{0.4cm} \een \nd Then, if
$\psi(\Lambda_2,...,\Lambda_{M})~$ is also a function of class
$C^2$ such that \ben \label{pde-s17}
\psi(\xi_1,...,\xi_{M-1})=c~,\hspace{1.5cm}
\left.\frac{\partial{\psi}}{\partial
\Lambda_k}\right|_{\xi_1,...,\xi_{M-1}}=d_{k-1}~,
\hspace{0,5cm}k=2,...,M. \een \nd exists a solution
$\mathcal{A}$ of (\ref{pde-s14}) in a neighborhood of
$\Lambda_1=a$ and $\Lambda_k=\xi_{k-1}$ that satisfies \ben
\label{pde-s18} \mathcal{A}(a,
\Lambda_2,\cdots,\Lambda_M)=\psi(\Lambda_2,\cdots,\Lambda_M) \een
\nd and is of class $C^2$.

\nd Regarding Cauchy-uniqueness,  it is known that if $F$
satisfies the Lipschitz condition \cite{pde-6}, \ben
\label{pde-s19}
\left|F\left(\Lambda_1,\cdots,\Lambda_M,{\mathcal{A}'},
\frac{\partial{\mathcal{A}'}}{\partial \Lambda_2},\cdots,
\frac{\partial{\mathcal{A}'}}{\partial \Lambda_M}\right)-
F\left(\Lambda_1,\cdots,\Lambda_M,{\mathcal{A}},
\frac{\partial{\mathcal{A}}}{\partial \Lambda_2},\cdots,
\frac{\partial{\mathcal{A}}}{\partial \Lambda_M}\right)\right|
~\leq \hspace{1.cm}\n\\
\leq K_1 \sum_{k=2}^M \left|
\frac{\partial{\mathcal{A}'}}{\partial \Lambda_k}-
\frac{\partial{\mathcal{A}}}{\partial \Lambda_k}\right| +K_2
\left|\mathcal{A}'-\mathcal{A}\right|~ \hspace{2.cm}K_1,~K_2~=
const.\hspace{1.cm} \een then, the solution of the initial value
problem for (\ref{pde-s14}) is unique. Note that in our case the
above condition  is verified always since the Legendre structure
the theory guarantee that

\ben \label{pde-s20}
F\left(\Lambda_1,\cdots,\Lambda_M,{\mathcal{A}},
\frac{\partial{\mathcal{A}}}{\partial \Lambda_2},\cdots,
\frac{\partial{\mathcal{A}}}{\partial \Lambda_M}\right)=
\frac{\partial{\mathcal{A}}}{\partial \Lambda_1}\propto
\frac{\partial{\alpha}}{\partial \lambda_1}= -\left\langle
x\right\rangle < \infty. \een

\section{10. Conclusions}

\nd On the basis of a variational principle based on Fisher
Information we have obtained in this paper a first order
differential equation for the Schr\"odinger energy-eigenvalues.
We have shown that the general solution exists and is unique.
This equation constitutes a necessary, but not sufficient
condition for $\alpha$ to be an energy-eigenvalue. Where does this
equation come from?

\nd It arises from the fact that the probability distribution that
minimizes Fisher's information measure $I$ (subject to
constraints) must be derived by solving a Schr\"odinger-like wave
equation, in which the normalization Lagrange multiplier $\alpha$
of the associated variational problem plays the role of an
energy-eigenvalue. A Legendre transform-invariant substructure
emerges then that inextricably links $I$ and $\alpha$ as Legendre
partners. This constitutes a new illustration of the power
of information-related tools in analizing physical problems.

\vspace{0.5cm}

\newpage
\appendix

\section{Appendix: FIM's translational transformation}

\nd The potential function \ben \label{ap-1}
U(x)=-~\frac{1}{8}\sum_{k=1}^M \lambda_k x^k. \n \een \nd can be
Taylor-expanded about $x=\xi$  \ben \label{ap-2} U(x)=\sum_{k=0}^M
\frac{U^{(k)}(\xi)}{k~!} (x-\xi)^k. \n \een

\nd The shift $u=x-\xi $  leads to \ben \label{ap-3}
\bar{U}(u)=~{U}(u+\xi)=~ \sum_{k=0}^{M}~\frac{U^{(k)}(\xi)}{k~!}
u^k,\een \nd which can be recast as \ben \label{ap-4}
\bar{U}(u)=-~\frac{1}{8} \sum_{k=0}^{M}~\lambda^{*}_k u^k ~,\een
\nd with \ben \label{ap-5} \lambda^{*}_k ~\equiv ~ -~ 8
~\frac{U^{(k)}(\xi)}{k~!}= -
~\frac{8}{k!}\sum_{j=1}^{M}~j(j-1)(j-2)\cdots(j-k+1)~\lambda_j~
{\xi}^{j-k}~. \een \nd The shifted-FIM corresponding to  $u=x-\xi
$ is obtained from (\ref{eq.1-12}) in the fashion [note that
$\langle ~ \rangle'$ indicates that the pertinent moment is
calculated with translation-transformed  (TF) eigenfunctions]

\ben \label{ap-6} I \, =\, -~4 \int {\psi} \frac{\partial^2
~}{\partial x^2} {\psi}~dx =\, -~4 \int \bar{\psi}
\frac{\partial^2 ~}{\partial u^2} \bar{\psi}~du =\, -~4
\left\langle \frac{\partial^2 ~}{\partial u^2} \right\rangle',
\een  where $\bar{\psi}=\bar{\psi}(u)$ is the TF of ${\psi}(x)$.
Now, using  the TF of (\ref{eq.1-5}) one easily finds
 \ben \label{ap-7} I\,=\, \int ~ \bar{\psi}_n \left(\alpha +
\sum_{k=0}^M~\lambda^{*}_k~u^k\right) \bar{\psi}_n~du ~, \een \nd
and one realizes that

 \ben \label{ap-8} I \,=\,\alpha
 + \sum_{k=0}^M~\lambda^{*}_k  \langle u^k\rangle' \,=\,\bar{\alpha}
 + \sum_{k=1}^M~\lambda^{*}_k  \langle u^k\rangle'~, \een
where \ben \label{ap-9}
\bar{\alpha}=\alpha+\lambda^{*}_0=\alpha-8U(\xi). \een

\nd Also, the virial theorem (\ref{virial-4}) leads to

\ben \label{ap-10}I= 4~\left\langle ~ \frac{\partial^2 ~}{\partial
u^2}\right\rangle' = -~4~\left\langle {u} ~ \frac{\partial
~}{\partial u} \bar{U}(u)\right\rangle' ~ =~-~
\sum_{k=1}^{M}~\frac{k}{2}~\lambda^{*}_k~ \langle {u}^{k}\rangle'.
\een

\nd The TF moments $\langle {u}^{k}\rangle'$ are related to the
original moments as

\ben \label{ap-11} \langle {u}^{k}\rangle' \, =\,  \int
{u}^{k}~\bar{\psi}^2(u) ~du \, =\,  \int {u}^{k}~{\psi}^2(u+\xi)
~du \,= \,  \int (x-\xi)^{k}~{\psi}^2(x) ~dx  \,= \,\langle
(x-\xi)^{k}\rangle.~\n  \een

\nd By recourse to the Newton-binomial  we write \ben
\label{ap-12} \int (x-\xi)^{k}~{\psi}^2(x) ~dx
~=~\sum_{j=1}^k{~(-1)^j~ \left( \begin{array}{c} k\\j  \end{array}
\right)~\xi^j~\int x^{k-j}~{\psi}^2(x) ~dx}, \een \nd and then we
finally have
    \ben \label{ap-13}
    \langle u^k \rangle' ~ = \,\langle (x-\xi)^{k}\rangle~ = ~\sum_{j=1}^k{~(-1)^j~
\left( \begin{array}{c} k\\j    \end{array} \right)~\xi^j~\langle
x^{k-j}\rangle}.\een

 \vskip 0.5cm

\noindent {\bf Acknowledgments-}
This work was partially supported by the programs
FQM-2445 and FQM-207 of the Junta de Andalucia-Spain,
and by CONICET (Argentine Agency).



\begin{thebibliography}{99}
\bibitem{virial} For any quantum system in stationary state,
 with a Hamiltonian does not involve time explicitly,
$H=- \frac{\hbar^2}{2m}~\vec{\nabla}~+~U(\vec{x}),~$the virial
theorem states  that \cite{greiner} $ \left\langle  -
\frac{\hbar^2}{m}\vec{\nabla} \right\rangle = \left\langle
\vec{x}.\vec{\nabla} U(\vec{x})\right\rangle,$ where the
expectation value is taken for stationary states of the
Hamiltonian.  
\end{thebibliography}
\end{document}